\documentclass[10pt,journal,compsoc]{IEEEtran}

\ifCLASSOPTIONcompsoc
  \usepackage[nocompress]{cite}
\else
  \usepackage{cite}
\fi

\ifCLASSINFOpdf
  \usepackage[pdftex]{graphicx}
\else
\fi

\usepackage{amsmath}

\ifCLASSOPTIONcompsoc
  \usepackage[caption=false,font=footnotesize,labelfont=sf,textfont=sf]{subfig}
\else
  \usepackage[caption=false,font=footnotesize]{subfig}
\fi

\usepackage[hidelinks]{hyperref}
\usepackage{booktabs}
\usepackage{multirow}
\usepackage{amssymb}
\usepackage{enumitem}
\usepackage{afterpage}

\usepackage{tikz}
\definecolor{mplred}{rgb}{1.0, 0.0039, 0.0980} 
\definecolor{mplblue}{rgb}{0.0, 0.0, 0.9765} 
\definecolor{mplgreen}{rgb}{0.0, 0.5059, 0.0941} 
\definecolor{mplpurple}{rgb}{0.5137, 0.0, 0.4902} 
\definecolor{mplorange}{rgb}{1.0, 0.6510, 0.1765} 

\begin{document}

\title{A Systematic Evaluation of Self-Supervised Learning for Label-Efficient Sleep Staging \\ with Wearable EEG}

\author{Emilio~Estevan$^*$,
        Mar\'{i}a~Sierra-Torralba,
        Eduardo~L\'{o}pez-Larraz,
        and~Luis~Montesano%
\IEEEcompsocitemizethanks{\IEEEcompsocthanksitem E. Estevan, M. Sierra-Torralba, E. L\'{o}pez-Larraz, and L. Montesano are with Bitbrain, Zaragoza, Spain. E-mail: \{emilio.estevan, maria.sierra, eduardo.lopez, luis.montesano\}@bitbrain.com%
\IEEEcompsocthanksitem L. Montesano is also with DIIS - I3A, Universidad de Zaragoza, Zaragoza, Spain. E-mail: \{montesano\}@unizar.es}%
\thanks{$^*$Corresponding author.}
}

\IEEEtitleabstractindextext{%
\begin{abstract}
Wearable EEG devices have emerged as a promising alternative to polysomnography (PSG). As affordable and scalable solutions, their widespread adoption results in the collection of massive volumes of unlabeled data that cannot be analyzed by clinicians at scale. Meanwhile, the recent success of deep learning for sleep scoring has relied on large annotated datasets. Self-supervised learning (SSL) offers an opportunity to bridge this gap, leveraging unlabeled signals to address label scarcity and reduce annotation effort. In this paper, we present the first systematic evaluation of SSL for sleep staging using wearable EEG. We introduce a structured benchmarking framework encompassing a range of SSL paradigms and propose a specialized pipeline tailored to the wearable EEG domain, evaluating them on two sleep databases acquired with the Ikon Sleep wearable headband: BOAS, a high-quality benchmark containing PSG and wearable EEG recordings with consensus labels, and HOGAR, a large collection of home-based, self-recorded, and unlabeled recordings. Three evaluation scenarios are defined to study label efficiency, representation quality, and cross-dataset generalization. Results show that SSL consistently improves classification performance by up to 10\% over supervised baselines, with gains particularly evident when labeled data is scarce. SSL achieves clinical-grade accuracy above 80\% leveraging only 5\% to 10\% of labeled data, while the supervised approach requires twice the labels. Additionally, the proposed domain-specific SSL pipeline outperforms the evaluated general-purpose EEG foundation models across all scenarios. Our findings demonstrate the potential of SSL to enable label-efficient sleep staging with wearable EEG, reducing reliance on manual annotations and advancing the development of affordable sleep monitoring systems.
\end{abstract}

\begin{IEEEkeywords}
Automatic sleep staging, electroencephalography (EEG), self-supervised learning, deep learning, wearable devices, label efficiency, representation learning.
\end{IEEEkeywords}}

\maketitle

\IEEEdisplaynontitleabstractindextext
\IEEEpeerreviewmaketitle


\IEEEraisesectionheading{\section{Introduction}\label{sec:introduction}}

\IEEEPARstart{S}{leep} plays a crucial role within the spectrum of human health, supporting a wide range of physiological processes \cite{besedovsky2019sleep, cappuccio2017sleep, harding2020sleep}. Sleep disorders affect a substantial portion of the population and represent a growing public health concern \cite{van2018sleep, latreille2018sleep}. Traditional sleep diagnostics have relied on full-night polysomnography (PSG), where sleep staging is performed manually by trained technicians to classify 30-second epochs into five sleep stages (Wake, N1, N2, N3, and REM) according to the guidelines of the American Academy of Sleep Medicine (AASM) \cite{Iber2007}. Despite its clinical value, PSG-based sleep scoring is resource-intensive, time-consuming, and subject to variability, with inter-scorer agreement typically between 80\% and 85\% \cite{danker2009interrater, rosenberg2013american, lee2022interrater}. EEG wearable devices have emerged as a promising alternative to conventional PSG, offering affordable, non-invasive, and home-based sleep monitoring \cite{de2024sleep}. Their growing popularity, driven by the prevalence of sleep disorders and the demand for accessible diagnostic tools, leads to the generation of massive volumes of EEG data that are impractical to label manually. While traditional PSG also faces this challenge, it is particularly pronounced in the context of wearable systems because of their scalability and ease of deployment.

Automatic sleep scoring powered by deep learning offers a scalable solution to the challenges introduced by wearable EEG technology. These methods enable accurate and efficient sleep staging at scale in complex environments \cite{zhang2024review}. State-of-the-art models can match or even surpass medical-grade performance in terms of inter-scorer agreement, even when trained on single-channel EEG data \cite{supratak2017deepsleepnet, supratak2020tinysleepnet, seo2020intra, eldele2021attention, esparza2024automatic}. Combined with wearable EEG, deep learning offers a path toward fully democratized access to sleep diagnostics by eliminating the need for manual scoring. Nevertheless, traditional deep learning methods are notoriously data-hungry, typically requiring large volumes of labeled data for effective training \cite{song2022learning}. In the context of wearable EEG, the acquisition of training data often involves parallel PSG recordings, where expert technicians annotate the PSG and transfer the labels to the wearable data. This is because experts rely on the rich, multimodal signals available in PSG (e.g., EOG, EMG, respiration), which are not present in wearable EEG, making standalone annotation unreliable without extensive retraining. As a result, the annotation of these datasets remains as resource-intensive as the manual scoring process, requiring up to 2 hours to label an 8-hour PSG recording \cite{vallat2021open}. This dependency on manual labeling significantly limits the scalability and deployment of conventional supervised deep learning approaches in both clinical and research applications.

Self-supervised learning (SSL) has the potential to address the challenges associated with supervised deep learning in EEG-based sleep staging using wearable technology. By leveraging the inherent structure of brain signals, SSL enables pre-training deep learning models on large volumes of unlabeled data, allowing the extraction of generalizable representations without the need for manual annotations \cite{banville2021uncovering}. Given that the mass adoption of wearable EEG devices will generate data at a scale that is not feasible to label manually for supervised model training, SSL provides a compelling strategy to incorporate this data into the learning pipeline and fully exploit its value. Thus, it can reduce dependence on expert annotations, improving label efficiency and making model development more affordable. It also enhances performance in low-label regimes by allowing models to be pre-trained on the full volume of collected data \cite{eldele2023self}, mitigating the data-hungry nature of deep learning. In addition, SSL improves generalization to real-world, home-based settings by reducing reliance on scorer-specific labels and minimizing the impact of annotation variability.

This study presents, to the best of our knowledge, the first systematic evaluation of self-supervised learning for automatic sleep staging in the context of wearable EEG devices. Our contributions are as follows:

\begin{itemize}
    \item We introduce a structured benchmarking framework to evaluate self-supervised learning techniques as a solution to leverage the large volumes of unlabeled data generated by wearable EEG, and propose a specialized SSL pipeline tailored to this domain, enabling the development of label-efficient systems that reduce dependence on manual annotation and outperform purely supervised baselines through the integration of unlabeled data.

    \item We provide practical insights into when and how to apply SSL in sleep staging by analyzing the trade-off between annotation effort and model performance, and by identifying key thresholds where SSL offers clear advantages over fully supervised learning.

    \item We validate an end-to-end sleep scoring pipeline based on affordable, wearable EEG devices and label-efficient algorithms, demonstrating its effectiveness in a realistic, home-based setting.
\end{itemize}

\section{Related Work}\label{sec:related-work}

\subsection{Automatic Sleep Staging with Deep Learning}

Deep learning has emerged as the state-of-the-art approach for automatic sleep staging, consistently outperforming traditional machine learning methods that rely on handcrafted features \cite{phan2022automatic}. Its capacity to learn directly from raw EEG data enables robust performance across large-scale datasets and challenging conditions, including noisy signals, reduced channel configurations, and inter-subject variability. Most existing deep learning models have been validated on well-established, expert-annotated PSG datasets such as SleepEDF \cite{goldberger2000physiobank, kemp2000analysis}, SHHS \cite{zhang2018national, quan1997sleep}, MASS \cite{o2014montreal}, ISRUC \cite{khalighi2016isruc}, or DREEM \cite{guillot2020dreem}. However, the dependence on large volumes of labeled data, typically generated through time-consuming manual annotation, can significantly limit the feasibility of deploying these models at scale in real-world or wearable settings.

Several notable deep learning architectures have driven substantial progress in performance within this domain. DeepSleepNet \cite{supratak2017deepsleepnet} combines dual CNN branches with different filter sizes to extract both temporal and frequency features, followed by bidirectional LSTMs, trained using a two-step procedure. XSleepNet \cite{phan2021xsleepnet} is a multi-view sequence-to-sequence model that jointly learns from raw EEG and time-frequency images by adaptively blending gradients from each view. AttnSleep \cite{eldele2021attention} introduces a modular, attention-based architecture combining a multi-resolution CNN, an adaptive recalibration module to model feature dependencies, and a temporal context encoder using multi-head self-attention with causal convolutions. SleepTransformer \cite{phan2022sleeptransformer} is a transformer-based, convolution- and recurrence-free model that processes time-frequency representations of EEG to enable interpretable and uncertainty-aware sleep staging.

\subsection{Self-Supervised Learning}

Self-supervised learning enables models to extract meaningful features from unlabeled data, making it a powerful strategy in domains where labeled data is scarce. SSL typically consists of two stages: a pretext task, which is designed to learn generalizable representations from the input data, and a downstream task, where the model is fine-tuned using labeled data to optimize performance on the target application \cite{rafiei2022self}. Among SSL approaches, contrastive learning methods, such as SimCLR \cite{chen2020simple}, BYOL \cite{grill2020bootstrap}, MoCo \cite{he2020momentum}, SimSiam \cite{chen2021exploring}, and Barlow Twins \cite{zbontar2021barlow}, learn representations by distinguishing between augmented views of the same input. Alternatively, masked prediction approaches, including BEiT \cite{bao2021beit}, Data2Vec \cite{baevski2022data2vec}, and MAE \cite{he2022masked}, train models to reconstruct masked input regions. Hybrid methods like CMAE \cite{huang2023contrastive} combine both strategies. 

In the context of SSL for EEG-based sleep staging, SimCLR and BYOL have been adapted from the image processing field by customizing their contrastive objectives to EEG signals in \cite{jiang2021self} and \cite{mai2021bootstrapnet}, respectively. ContraWR \cite{yang2021self} employs global representations across the dataset to guide contrastive learning. TS-TCC \cite{eldele2021time} combines efficient data augmentations with both temporal and contextual contrastive components. CoSleep \cite{ye2021cosleep} learns generalizable representations through a co-training scheme from multiple views to mine positive samples, along with a queue and a momentum encoder to build a memory bank of negative samples. BENDR \cite{kostas2021bendr} leverages contrastive learning by comparing reconstructed features, produced by a Transformer with masked inputs, to original features extracted by a preceding CNN encoder. MAEEG \cite{chien2022maeeg} extends BENDR by introducing additional mapping layers to enable reconstruction loss. mulEEG \cite{kumar2022muleeg} utilizes multiple views of EEG and combines diversity and contrastive losses. NeuroNet \cite{lee2024neuronet} adopts a hybrid approach by integrating contrastive learning and masked prediction tasks using a Transformer autoencoder on top of a CNN encoder.

While these earlier methods typically pre-train on relatively restricted, homogeneous datasets for specific applications, the SSL paradigm within EEG research has recently shifted toward the development of foundation models \cite{lai2025simple}. These large-scale models are designed to learn hierarchical representations from massive, heterogeneous unlabeled EEG datasets, yielding general-purpose representations that transfer to diverse downstream tasks with minimal adaptation \cite{liu2026eeg}. Architecturally, the majority of these models rely on Transformer-based backbones \cite{liu2026eeg}, whereas others incorporate novel state-space models (SSMs) such as Mamba \cite{wang2025eegmamba, tegon2025femba}. The most common pre-training objective is masked signal modeling \cite{lee2025comprehensive}, where models are tasked with reconstructing the original signal \cite{zhang2023brant, wang2024eegpt, wang2024cbramod, zhou2025csbrain, ouahidi2025reve}, patch embeddings \cite{yang2023biot}, or discrete codebook tokens \cite{jiang2024large, barmpas2025neurorvq} from masked positions. Alternative strategies leverage autoregressive training \cite{cui2024neuro, jiang2024neurolm} or contrastive learning frameworks \cite{ahuja2025ss, wang2025lead}. Within the sleep domain, this paradigm primarily centers on pre-training large-scale polysomnographic models. Recent studies leverage massive, multi-center cohorts to address the inherent variability of sensor configurations \cite{thapa2026multimodal, deng2025unified, fox2025foundational}. Transformer backbones dominate for encoding long-context physiological data \cite{fox2025foundational, yu2025language}, while alternative approaches explore deep generative frameworks \cite{van2025deep}. Pre-training objectives include cross-modal contrastive learning \cite{thapa2026multimodal}, domain-adaptive pre-training \cite{deng2025unified}, and autoregressive sequence modeling \cite{yu2025language}. These SSL approaches uncover latent sleep dynamics that bypass the temporal constraints and subjectivity of traditional AASM scoring \cite{coon2025foundation}, demonstrating remarkable transferability in predicting various health-related outcomes \cite{thapa2026multimodal, deng2025unified, yu2025language}.

\subsection{Sleep Scoring with Wearable EEG Devices}

Recent research has increasingly focused on evaluating wearable EEG devices for sleep staging, primarily by benchmarking their performance against the gold standard of PSG. A systematic review by De Gans et al. \cite{de2024sleep} analyzed 60 studies encompassing 34 unique EEG-based wearables, highlighting a surge in interest, with over half of the reviewed publications appearing from 2020 onward. The reviewed devices demonstrated promising accuracy in sleep staging, often approaching that of PSG. However, the vast majority of studies were conducted in controlled, clinical settings, with only 12\% of them validating wearables in home environments. Moreover, most approaches relied on fully labeled datasets, leaving the potential of label-efficient training paradigms, such as SSL, largely unexplored. Khalique et al. \cite{khalique2025label} provided an important first step toward integrating SSL with portable EEG for in-home sleep scoring. Nonetheless, the scope of their work is limited to a single SSL method and training setting. Our study conducts the first comprehensive evaluation of SSL for wearable EEG sleep staging, comparing multiple state-of-the-art SSL methods across different evaluation scenarios using both clinically annotated and large-scale unlabeled home recordings to analyze label efficiency, representation quality, and cross-dataset generalization.

Across the main categories of wearable EEG systems, headband devices like the Dreem headband have shown strong agreement with PSG, achieving 83.5\% accuracy and 83.8\% F1 score for automatic sleep staging against expert consensus \cite{arnal2020dreem}. Hsieh et al. proposed an eye mask-based system that integrates EEG and EOG sensing with mobile deep learning, obtaining over 86\% overall agreement with manual PSG annotations in four-class sleep scoring \cite{hsieh2021home}. The in-ear EEG sensor developed in \cite{nakamura2019hearables} enabled home-based five-stage automatic sleep classification with 74.1\% accuracy and a Cohen's kappa of 0.61 versus PSG. Validated in a clinical sample, the UMindSleep single-channel EEG forehead device achieved sensitivities above 79\%, specificities over 83\%, and kappa agreements ranging from 0.69 to 0.79 when compared to PSG in four-class sleep staging \cite{chen2023validation}. The WPSG-I portable PSG system demonstrated high agreement with standard PSG, with 95.8\% accuracy ($\kappa$ = 0.92) for manual scoring and 89.7\% ($\kappa$ = 0.80) for automated staging \cite{li2025exploring}. Finally, the HARU patch-type forehead EEG system, using a multi-sensor sheet and deep learning-based staging, reached 78.6\% accuracy and a macro F1 score of 73.4\%, showing performance comparable to clinical PSG in healthy participants \cite{matsumori2022haru}.

\section{Methods}\label{sec:methods}

\subsection{Model Architecture}\label{sec:model_arq}

\begin{figure*}[!t]
\centering
\includegraphics[width=0.95\linewidth]{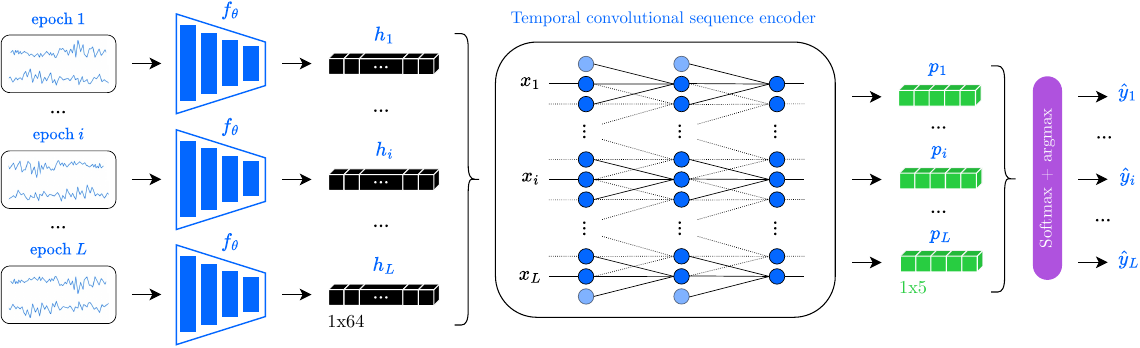}
\vspace{-0.2cm}
\caption{Overview of the deep learning model architecture for automatic sleep staging. Architectural details are available in \cite{esparza2024automatic}.}
\label{fig:model_arq}
\vspace{-0.25cm}
\end{figure*}

The deep learning model architecture, previously validated in our earlier studies \cite{esparza2021sleepbci, esparza2024automatic} and depicted in Fig. \ref{fig:model_arq}, follows a sequence-to-sequence framework commonly used in sleep staging \cite{phan2022automatic}. The model receives as input a window of $L$ raw EEG epochs $(x_1, \dots, x_L)$, where each $x_i \in \mathbb{R}^{C \times T}$ represents a 30-second segment of EEG data with $C$ channels and $T$ time points, for $1 \leq i \leq L$. The output is a sequence of predicted sleep stage labels $(\hat{y}_1, \dots, \hat{y}_L)$. More specifically, the network consists of two main components. Firstly, an epoch encoder sub-model $f_{\theta}$ processes each input epoch independently to produce a feature vector $h_i \in \mathbb{R}^{64}$. This encoder follows a 1D convolutional architecture designed to extract robust, temporal-invariant features that capture the intrinsic properties of each EEG epoch, regardless of when they occur in the night (i.e., intra-epoch modeling). On the other hand, a temporal convolutional sequence encoder processes the sequence of encoded epochs $(h_1, \dots, h_L)$ to generate the corresponding sequence of logit vectors $(p_1, \dots, p_L)$. Its core function is to model temporal dependencies across successive epochs (i.e., inter-epoch modeling), enabling a sequence-to-sequence sleep stage classification by imitating the approach carried by technicians during manual scoring. Finally, softmax and argmax operations are applied to the output logits to determine the sleep labels.

\subsection{Self-Supervised Learning Techniques}\label{sec:ssl_methods}

\begin{figure*}[!t]
\centering
\includegraphics[width=1\linewidth]{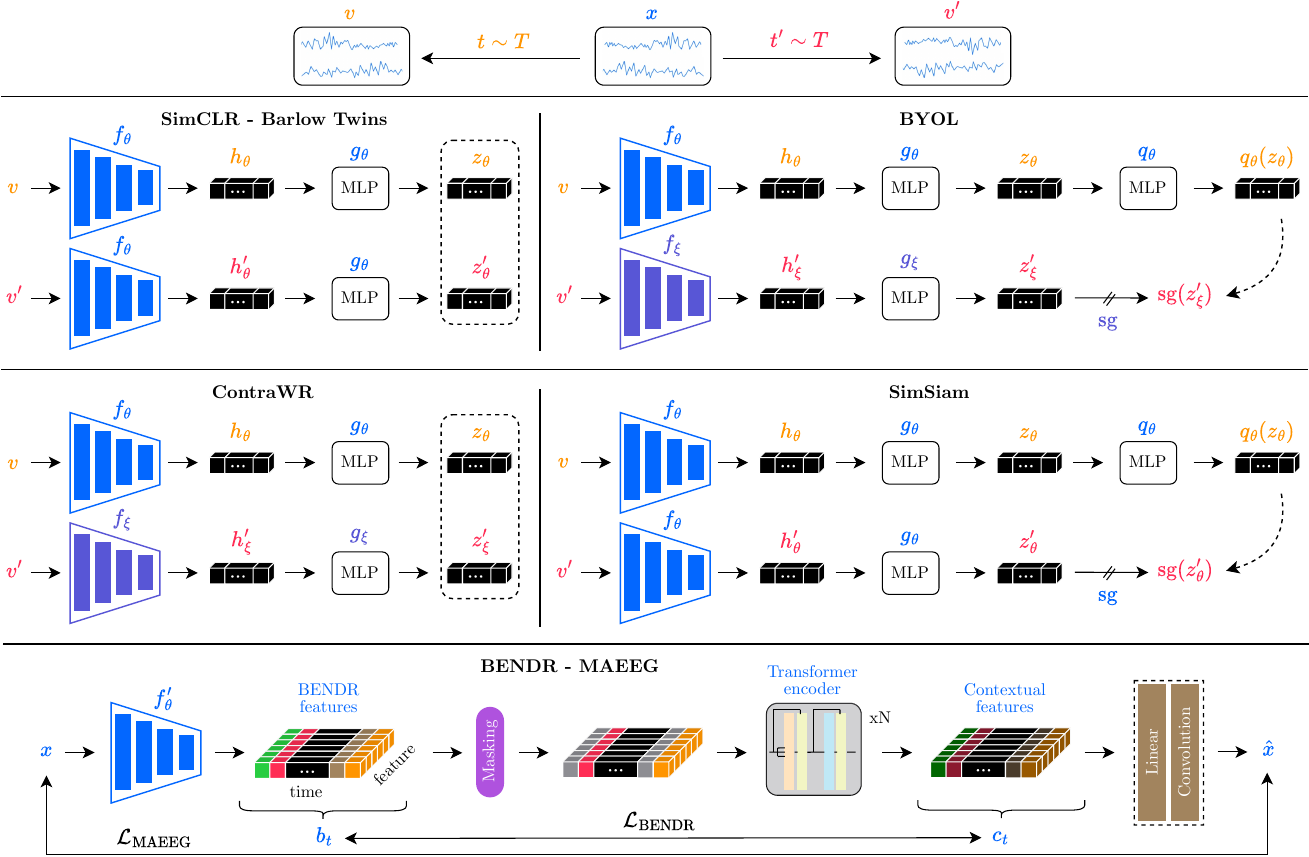}
\vspace{-0.55cm}
\caption{Illustration of the self-supervised learning strategies used for model pre-training on unlabeled EEG data.}
\label{fig:ssl_methods}
\vspace{-0.25cm}
\end{figure*}

In this section, we describe the set of SSL methods employed to pre-train our epoch encoder $f_{\theta}$ using unlabeled EEG data. The selection focused on well-established approaches in the literature, particularly in computer vision \cite{gui2024survey, chen2024progress} and, more recently, in the time-series domain \cite{eldele2023self, rafiei2022self, lee2024neuronet}. These paradigms were originally established over the past few years, and they remain the cornerstone pre-training objectives in current SSL research. Crucially, they form the algorithmic backbone of the most recent EEG foundation models \cite{lai2025simple, liu2026eeg, lee2025comprehensive}, which continue to rely on masked signal modeling \cite{wang2024cbramod, zhou2025csbrain, ouahidi2025reve} and contrastive learning \cite{ahuja2025ss, wang2025lead, thapa2026multimodal}. Consequently, this selection provides a comprehensive and representative framework for evaluating SSL capabilities within the scope of this study. An overview of these techniques is provided in Fig. \ref{fig:ssl_methods}.

\textbf{SimCLR} \cite{chen2020simple}: Given an input sample $x$, two data augmentations $t \sim T$ and $t' \sim T$ are applied to generate a positive pair $v = t(x)$ and $v' = t'(x)$. These augmented views pass through a neural encoder $f_{\theta}$ (\textit{representation head}) to generate representations $h_{\theta} = f_{\theta}(v)$ and $h_{\theta}' = f_{\theta}(v')$. A lightweight \textit{projection head} $g_{\theta}$ maps the representations to the latent space $z_{\theta} = g_{\theta}(h_{\theta}) = W^{(2)}\sigma(W^{(1)}h_{\theta})$, being $\sigma$ a ReLU non-linearity, and $z_{\theta}' = g_{\theta}(h_{\theta}')$, where the \textit{NT-Xent} contrastive loss is applied after a final batch normalization.

\textbf{BYOL} \cite{grill2020bootstrap}: Employs two asymmetric networks (\textit{online} and \textit{target}) to learn representations without negative samples. Given $v$ and $v'$, the online network, parameterized by $\theta$, includes a representation encoder $f_{\theta}$, a projection head $g_{\theta}$, and an additional \textit{prediction head} $q_{\theta}$, outputting $q_{\theta}(z_{\theta})$. The target network shares the same architecture as the online network, excluding the prediction head, and is parameterized by $\xi$, an exponential moving average (EMA) of $\theta$, producing $z_{\xi}'$. The loss function minimizes the mean squared error with respect to $\theta$ but not $\xi$ (stop-gradient $\text{sg}(z_{\xi}')$) between the $\ell_2$-normalized predictions and target projections $\overline{q_{\theta}}(z_{\theta})$ and $\overline{z}_{\xi}'$, and is symmetrized by separately feeding $v'$ to the online network and $v$ to the target network.

\textbf{SimSiam} \cite{chen2021exploring}: This technique follows the same pipeline as BYOL but shares the same set of weights $\theta$ across both branches. Compared to previous approaches, $g_{\theta}$ adopts a deeper architecture by including one more fully-connected layer. The loss minimizes the negative cosine similarity between the outputs $q_\theta(z_\theta)$ and $z_{\theta}'$, applying a stop-gradient operation to the second branch, and is symmetrized and averaged over all samples in the minibatch.

\textbf{Barlow Twins} \cite{zbontar2021barlow}: The method adopts the same architecture as SimCLR, with the exception of a deeper $g_{\theta}$ (as SimSiam). Its innovative loss measures the cross-correlation matrix $C$ between the outputs of the two branches fed with distorted views of a batch of samples, and tries to make it close to the identity. The invariance term of the loss makes the embeddings invariant to the augmentations by driving the diagonal of $C$ to 1, while the redundancy reduction term aims to decorrelate the vector components by pushing the off-diagonal elements of $C$ toward 0.

\textbf{ContraWR} \cite{yang2021self}: Building on the symmetric pipeline of SimCLR, but incorporating an additional EMA network for the second branch, as BYOL, ContraWR introduces a triplet loss function that enforces greater similarity, based on a Gaussian kernel, between a positive pair $(z_{{\theta}}, z'_{\xi})$ than between $z_{{\theta}}$ and a world representation $z_w$. This world representation can be estimated either globally, as the average embedding across the dataset, or in an instance-aware manner, specific to each $z_{{\theta}}$. In both cases, $z_w$ is approximated using a Monte Carlo estimation within each minibatch.

\textbf{BENDR} \cite{kostas2021bendr}: The first stage employs a convolutional encoder $f_{\theta}'$ that downsamples the input signal $x$ into a sequence of latent vectors (BENDR) with time and feature dimensions. Then, a masking operation is applied along the time dimension, and a subsequent Transformer encoder is tasked to reconstruct the masked data in the latent space. This reconstruction is guided by the NT-Xent contrastive loss, computed between the BENDR features (CNN encoder output) and the contextual features (Transformer output).

\textbf{MAEEG} \cite{chien2022maeeg}: This framework extends the BENDR architecture by adding two layers that project the contextual features back to the original EEG input space, enabling both temporal and spatial reconstruction. By doing so, MAEEG learns representations by minimizing a reconstruction loss between the raw input EEG $x$ and its reconstruction $\hat{x}$, directly in the signal domain, without contrastive learning.

Once pre-training with unlabeled data is completed, the backbone network $f_{\theta}$ is employed for downstream tasks. The notation $f_{\theta}'$, used in BENDR and MAEEG, refers to the portion of the encoder network up to the last convolutional layer, excluding the final temporal pooling operation that yields the 1D feature vector, representing a minor architectural difference. This preserves the time and channel dimensions, consistent with the 2D output requirements of these methods for effective SSL pre-training. During downstream evaluation, the convolutional layers are initialized with the pre-trained weights and the pooling operation is applied, resulting in the exact same downstream architecture, independent of the pre-training technique. Appendix \ref{app:data-aug} provides details on the two data augmentation sets $T_1$ and $T_2$ used in this work. Given a transformation set $T \in \{T_1, T_2\}$, a random augmentation $t \sim T$ is randomly selected with equal probability and applied independently to every EEG channel within each sleep epoch. For SimCLR, SimSiam, Barlow Twins, and ContraWR, we adopt the first set $T_1$, while BYOL relies on the second set $T_2$, as it resulted in better performance.

\section{Experiments}\label{sec:experiments}

\subsection{Device and Datasets}

We employ sleep EEG data recorded with the Ikon Sleep\footnote{\url{https://www.bitbrain.com/neurotechnology-products/textile-eeg/ikon-sleep}} wearable headband (Bitbrain, Zaragoza, Spain) to evaluate the automatic sleep staging framework. The device records two frontal EEG electrodes (AF7 and AF8), sampled at 256 Hz, with reference and ground placed on the left and right mastoids, respectively. It also features a photoplethysmography (PPG) sensor and an inertial measurement unit (IMU). Two novel datasets acquired with this device are used in the present study.

\textbf{BOAS} \cite{boas_dataset}: The Bitbrain Open Access Sleep (BOAS) dataset includes 128 overnight recordings from healthy adults. Each participant was simultaneously monitored using a clinical-grade PSG system and the Ikon Sleep wearable EEG headband. PSG data consists of EEG (F3, F4, C3, C4, O1, O2), EOG, EMG, PPG, breathing activity, and respiratory airflow. Sleep stages were labeled following AASM guidelines \cite{Iber2007} by three independent experts. A consensus label was assigned when at least two scorers agreed, with disagreements resolved by a fourth scorer. Labels correspond to: Wake (0), N1 (1), N2 (2), N3 (3), REM (4), disconnections (8), and artifacts (-2). In this study, only annotations 0 to 4 were retained. We utilize the frontal EEG channels (AF7 and AF8) recorded by the wearable headband.

\textbf{HOGAR} \cite{hogar_dataset}: This database is part of an ongoing research project aimed at enabling early detection and treatment of cognitive decline through large-scale, automatic, and accessible analysis of EEG recorded during sleep and wakefulness. As of November 2024, it includes 239 overnight recordings from elderly participants (aged \textgreater\! 60 years) collected in home environments using the Ikon Sleep wearable EEG headband. This dataset contains unlabeled EEG data in the context of sleep staging, as PSG is not available for conventional labeling, collected under wearable, real-world conditions, where users fully self-administer the device configuration. As in BOAS, only the frontal EEG channels (AF7 and AF8) are selected. 

Therefore, the HOGAR dataset provides a substantial volume of unlabeled EEG data suitable for representation learning, as a direct consequence of adoption of wearable EEG devices at scale, while the BOAS database is used as a high-quality reference collection for supervised training and validation against gold-standard PSG annotations within the sleep classification task. This dual-dataset configuration establishes a robust framework for evaluating SSL in wearable sleep staging, especially given the limitations of currently available data in this domain. As highlighted by recent reviews \cite{de2024sleep}, the literature on consumer EEG wearables is constrained by a scarcity of published validation studies, small sample sizes, and inherent data heterogeneity. These factors complicate cross-dataset comparisons and hinder large-scale analyses. Consequently, pairing a home-based, real-world unlabeled dataset (HOGAR) with a clinical, PSG-validated benchmark (BOAS) constitutes a comprehensive and reliable approach for validating the proposed SSL methodology within the scope of this study.

\subsection{Evaluation Pipeline}\label{sec:evaluation}

This section describes the evaluation pipeline employed in this work, illustrated in Fig. \ref{fig:pipeline}. Each recording is segmented into 30-second epochs, downsampled from 256 Hz to 128 Hz, and filtered using a band-pass filter between 0.5 and 45 Hz. Subsequently, the pretext task begins with a Z-score normalization of the data, followed by the SSL stage that trains the epoch encoder $f_{\theta}$ using one of the SSL methods introduced in Section \ref{sec:ssl_methods}. This part aims to learn general feature representations from unlabeled EEG signals, providing better-than-random initial parameters that boost model performance in later tasks. Training and hyperparameter configurations for each method are detailed in Appendix \ref{app:ssl-config}. Beyond numerical insights, the quality of the feature representations is additionally examined using UMAP \cite{mcinnes2018umap}.

\afterpage{
\begin{figure*}[!t]
\centering
\includegraphics[width=0.95\linewidth]{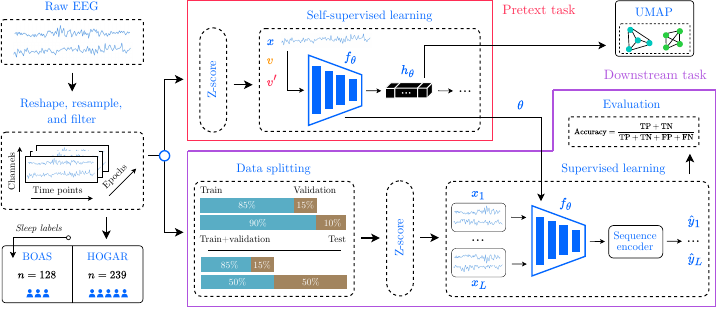}
\vspace{-0.15cm}
\caption{Evaluation pipeline for the automatic sleep staging framework. First, the raw EEG data is preprocessed and partitioned to feed the subsequent stages. Then, the pretext task consisting of self-supervised learning extracts general representations from unlabeled signals. These representations are examined using UMAP. Finally, the downstream task involves supervised training with labeled EEG data for sleep classification.}
\label{fig:pipeline}
\vspace{-0.25cm}
\end{figure*}
}

The downstream task consists of supervised sleep stage classification. Labeled EEG data is split into training, validation, and test sets, with a Z-score normalization using only the training data. In this stage, the epoch encoder $f_{\theta}$ can be initialized with the weights $\theta$ learned during the pretext task, providing a more informed starting point. The sequence encoder is initialized from scratch. This model is fed with windows of size $L$ = 100 and trained for 100 epochs using the Adam optimizer $(lr = 0.0001,\ \beta_1 = 0.9,\ \beta_2 = 0.999,\ \epsilon = 1^{-8})$, which minimizes the cross-entropy loss. The accuracy metric is employed as the primary performance indicator, consistent with  \cite{yang2021self, eldele2021time, kumar2022muleeg, ye2021cosleep}.

Two main evaluation methodologies are commonly used in SSL \cite{chen2020simple}: semi-supervised learning, which involves fine-tuning a pre-trained model (with unlabeled data) on a labeled dataset, and linear evaluation, where a linear classifier is trained on top of frozen representations to assess their quality. In this work, we adopt the semi-supervised setting, as our goal is to evaluate the full potential of SSL in a realistic sleep staging scenario, aligning more closely with the practical application and potential deployment of SSL to real-world problems. To evaluate the proposed pipeline under a semi-supervised framework, we define three experimental scenarios that enable a systematic analysis of label efficiency, cross-dataset generalization, and the overall effectiveness of SSL in wearable EEG-based sleep staging.

\textbf{Scenario 1 -- SSL on HOGAR with Supervised Cross-Validation on BOAS}: The backbone network $f_{\theta}$ is first pre-trained using SSL on the entire unlabeled HOGAR dataset. The learned weights are then transferred to initialize the model for supervised training on a selected data partition of the BOAS dataset, using a recording-wise 10-fold cross-validation. Within each fold, the training subset is further split into 85\% for training and 15\% for validation. Every labeled BOAS partition is randomly selected $N$ times, and results are averaged across executions. This proportion of labeled data used for supervised training is progressively increased to evaluate the framework under different data availability conditions, and to compare performance against a purely supervised baseline without SSL pre-training.

\textbf{Scenario 2 -- SSL on HOGAR with Supervised Evaluation on a Fixed BOAS Test Set}: The same SSL pre-training is performed on the unlabeled EEG signals from the HOGAR dataset. In contrast, the BOAS dataset is split into two parts: a large, constant test set comprising 50\% of the data, and a remaining portion used for training (85\%) and validation (15\%). As before, recordings from the training set are incrementally selected, while the test set remains fixed. Thus, this setup provides a more consistent and representative test set, allowing fair and controlled comparison across distinct training sizes. The test partition is randomly selected $N$ times, and for each split, $M$ training sets are chosen.

\textbf{Scenario 3 -- SSL and Supervised Cross-Validation Within BOAS}: Both SSL pre-training and supervised training are performed entirely on the BOAS dataset. A portion of the available recordings is selected for SSL, while the remaining data is used for supervised training using a 10-fold cross-validation. This configuration enables the evaluation of the learned representations within the same dataset, allowing direct comparison with the general features extracted from the HOGAR dataset. The scenario is also repeated $N$ times with different random data partitions.

In addition to the purely supervised baseline, we incorporate two leading, high-performing EEG foundation models, LaBraM \cite{jiang2024large} and CBraMod \cite{wang2024cbramod}, alongside SleepFM \cite{thapa2026multimodal}, pre-trained on the largest multimodal sleep corpus to date, for scenarios 1 and 2. This inclusion allows us to benchmark our domain-specific SSL backbone against the large-scale, general-purpose representations that define the current frontier of SSL in EEG research. Given that the emergence of publicly available foundation models challenges the necessity of specialized SSL pipelines, empirically determining whether these massive architectures outperform our domain-focused pre-training is a critical aspect of this study. Furthermore, this paradigm raises a practical question for real-world deployment: whether such general-purpose representations can adequately replace domain-specific pre-training on locally acquired sleep data. This comparison is clinically relevant because wearable monitoring systems are often deployed in specific populations and acquisition settings that differ from the vast, heterogeneous datasets used to train massive foundation models. Although a comprehensive benchmarking of specialized SSL against foundation models remains out of scope for this work, evaluating both approaches provides valuable insights into which strategy could be more appropriate for data-limited, context-specific sleep staging applications, such as those modeled in our proposed scenarios. Details regarding the configuration and fine-tuning of these models are provided in Appendix \ref{app:fm-config}.

\section{Results}\label{sec:results}

\subsection{Performance of SSL in Different Data Settings}

\subsubsection{Scenario 1: SSL on HOGAR with Supervised Cross-Validation on BOAS}

The first row of Table \ref{tab:res-HOGAR-std3-cv} corresponds to a purely supervised baseline, while the remaining ones include models pre-trained on large-scale external corpora (three foundation models) and our proposed SSL pipeline pre-trained on the entire HOGAR dataset, respectively. SSL methods consistently outperform the supervised baseline across nearly all label regimes. Notably, Barlow Twins achieves an improvement of +8.08\% with only 7.5\% labeled data, while SimCLR offers gains of +2.98\% at 20\% and +0.78\% with the full database. The consistent improvements support the capability of SSL approaches to learn generalizable representations from the HOGAR dataset that are highly useful for the downstream task of sleep staging performed on the BOAS dataset. These gains diminish as more labeled data becomes available, highlighting the effectiveness of SSL in low-label settings. Regarding the foundation models, LaBraM and CBraMod outperform SleepFM but only significantly surpass the supervised baseline in the most restricted data regime, offering gains of +2.84\% and +1.89\%, respectively. Beyond this setting, their performance becomes comparable to or even worse than the supervised approach. LaBraM outperforms CBraMod in low-data scenarios (74.95\% vs. 74.00\% with 7.5\% data), whereas CBraMod scales better with abundant data (86.08\% vs. 84.76\% with 100\% data). Crucially, our domain-specific SSL pipeline consistently outperforms these massive architectures across all settings, demonstrating that targeted pre-training yields superior task-specific representations. Specifically, Barlow Twins exceeds LaBraM by +5.24\% at 7.5\% labeled data, while SimCLR surpasses CBraMod by +2.47\% at 20\% and +1.13\% with the full dataset. Accuracy variability is higher with low-data values but stabilizes when increasing the number of labeled samples. Among all methods, SimCLR, Barlow Twins, and ContraWR emerge as the best-performing SSL techniques. The final column represents the definitive classification performance of the system using all available EEG data, with SimCLR achieving an accuracy of 87.21\%.

\begin{table}[!t]
\renewcommand{\arraystretch}{1.3}
\setlength\tabcolsep{3.9pt}
\centering
\caption{Accuracy and standard deviation results for evaluation scenario 1: SSL on HOGAR with Supervised Cross-Validation on BOAS.}
\vspace{-0.25cm}
\scriptsize
\begin{tabular}{@{}lccccc@{}}
\toprule
\multirow{2}{*}{} & \multicolumn{5}{c}{Percentage of labeled data} \\ \cmidrule(l){2-6} 
 & 7.5 & 15 & 20 & 60 & 100 \\ \midrule
\textit{Supervised} & 72.11$\pm$5.97 & 80.35$\pm$2.33 & 81.34$\pm$1.49 & 84.97$\pm$1.06 & 86.43$\pm$0.05 \\ \midrule
\textit{LaBraM-Base} & 74.95$\pm$6.22 & 79.95$\pm$1.39 & 81.24$\pm$1.53 & 83.67$\pm$1.19 & 84.76$\pm$0.12 \\
\textit{CBraMod} & 74.00$\pm$4.35 & 78.81$\pm$1.89 & 81.85$\pm$1.34 & 84.48$\pm$0.40 & 86.08$\pm$0.01 \\
\textit{SleepFM} & 66.30$\pm$2.88 & 72.82$\pm$4.13 & 75.66$\pm$3.87 & 79.98$\pm$1.91 & 82.08$\pm$0.99 \\ \midrule
\textit{SimCLR} & 79.22$\pm$4.86 & 83.15$\pm$1.63 & \textbf{84.32$\pm$0.55} & 85.49$\pm$1.14 & \textbf{87.21$\pm$0.15} \\
\textit{BYOL} & 74.60$\pm$6.24 & 80.73$\pm$1.66 & 83.25$\pm$1.30 & 85.18$\pm$0.62 & 87.17$\pm$0.30 \\
\textit{SimSiam} & 76.98$\pm$5.25 & 81.96$\pm$1.61 & 83.66$\pm$1.00 & 85.50$\pm$0.92 & 86.56$\pm$0.38 \\
\textit{Barlow Twins} & \textbf{80.19$\pm$3.97} & 82.91$\pm$1.83 & 84.14$\pm$1.34 & \textbf{85.73$\pm$1.19} & 87.08$\pm$0.14 \\
\textit{ContraWR} & 79.38$\pm$4.91 & \textbf{83.37$\pm$1.53} & 84.09$\pm$1.08 & 85.37$\pm$1.33 & 87.04$\pm$0.19 \\
\textit{BENDR} & 79.95$\pm$2.96 & 81.55$\pm$0.71 & 83.34$\pm$0.76 & 85.38$\pm$0.75 & 86.48$\pm$0.03 \\
\textit{MAEEG} & 75.87$\pm$4.95 & 80.00$\pm$0.74 & 82.13$\pm$1.64 & 85.01$\pm$1.19 & 86.26$\pm$0.09 \\ \bottomrule
\end{tabular}

\label{tab:res-HOGAR-std3-cv}
\vspace{-0.25cm}
\end{table}

\subsubsection{Scenario 2: SSL on HOGAR with Supervised Evaluation on a Fixed BOAS Test Set}

\begin{table*}[!t]
\renewcommand{\arraystretch}{1.3}
\centering
\caption{Accuracy and standard deviation results for evaluation scenario 2: SSL on HOGAR with Supervised Evaluation on a Fixed BOAS Test Set.}
\vspace{-0.25cm}
\scriptsize
\begin{tabular}{@{}lcccccccc@{}}
\toprule
\multirow{2}{*}{} & \multicolumn{8}{c}{Percentage of labeled data} \\ \cmidrule(l){2-9} 
 & 2.5 & 5 & 10 & 20 & 40 & 60 & 80 & 100 \\ \midrule
\textit{Supervised} & 63.35$\pm$2.14 & 69.42$\pm0.59$ & 74.46$\pm$1.01 & 78.64$\pm$0.75 & 82.46$\pm$0.51 & 84.14$\pm$0.70 & 84.47$\pm$0.61 & 85.02$\pm$0.90 \\ \midrule
\textit{LaBraM-Base} & 63.43$\pm$0.88 & 68.45$\pm$1.89 & 72.66$\pm$0.91 & 77.09$\pm$0.68 & 80.74$\pm$0.53 & 81.83$\pm$0.50 & 82.56$\pm$0.71 & 83.33$\pm$0.66 \\
\textit{CBraMod} & 63.52$\pm$0.41 & 68.25$\pm$0.95 & 73.19$\pm$1.06 & 76.99$\pm$1.15 & 81.12$\pm$1.07 & 82.51$\pm$0.64 & 83.43$\pm$0.76 & 84.39$\pm$1.01 \\
\textit{SleepFM} & 64.10$\pm$1.76 & 66.32$\pm$4.11 & 66.41$\pm$2.78 & 68.80$\pm$2.42 & 76.07$\pm$1.87 & 76.67$\pm$2.94 & 79.08$\pm$1.79 & 80.43$\pm$1.09 \\ \midrule
\textit{SimCLR} & \textbf{70.10$\pm$0.61} & \textbf{76.15$\pm$2.26} & \textbf{79.78$\pm$1.59} & \textbf{82.48$\pm$0.96} & \textbf{83.95$\pm$0.65} & \textbf{85.19$\pm$0.71} & 85.36$\pm$0.58 & 86.05$\pm$1.31 \\
\textit{BYOL} & 65.48$\pm$0.83 & 72.49$\pm$2.41 & 75.92$\pm$0.77 & 79.81$\pm$1.22 & 83.17$\pm$0.80 & 84.64$\pm$0.98 & 85.10$\pm$1.06 & 85.92$\pm$1.08 \\
\textit{SimSiam} & 65.45$\pm$1.27 & 73.07$\pm$1.48 & 77.49$\pm$0.84 & 80.53$\pm$0.99 & 83.35$\pm$0.27 & 84.30$\pm$1.06 & 84.85$\pm$0.71 & 85.65$\pm$1.07 \\
\textit{Barlow Twins} & 68.96$\pm$1.93 & 75.34$\pm$2.14 & 79.04$\pm$1.38 & 82.37$\pm$0.97 & 83.92$\pm$0.55 & 85.03$\pm$0.53 & \textbf{85.55$\pm$0.85} & \textbf{86.40$\pm$0.80} \\
\textit{ContraWR} & 67.38$\pm$1.71 & 74.61$\pm$2.65 & 78.40$\pm$1.95 & 81.10$\pm$0.92 & 83.92$\pm$0.30 & 84.84$\pm$0.74 & 85.37$\pm$0.78 & 85.77$\pm$0.73 \\
\textit{BENDR} & 66.80$\pm$1.68 & 73.94$\pm$1.80 & 78.11$\pm$1.33 & 80.79$\pm$1.13 & 83.49$\pm$0.71 & 84.46$\pm$0.74 & 84.62$\pm$0.72 & 85.52$\pm$0.46 \\ 
\textit{MAEEG} & 65.14$\pm$0.93 & 71.43$\pm$1.68 & 74.82$\pm$1.33 & 78.74$\pm$0.34 & 82.24$\pm$0.78 & 83.61$\pm$0.79 & 84.62$\pm$0.62 & 85.09$\pm$0.98 \\ \bottomrule
\end{tabular}

\label{tab:res-HOGAR-std3}
\vspace{-0.2cm}
\end{table*}

Table \ref{tab:res-HOGAR-std3} reports the results for evaluation scenario 2, where the percentages of labeled data indicate the portion selected for training and validation from the 50\% of the BOAS dataset. In this setting, variability is generally lower, especially in low-data regimes, due to the constant test set. As in scenario 1, nearly all SSL methods outperform the purely supervised baseline across label availability conditions. SimCLR shows a notable improvement of +6.75\% with only 2.5\% labeled data, and +3.84\% with 20\%. Barlow Twins achieves a +1.38\% gain over the baseline with the full training set. These improvements, while slightly more moderate than in scenario 1, again demonstrate the benefits of SSL in low-label regimes. SimCLR and Barlow Twins emerge as the most effective methods, whereas MAEEG shows consistently lower performance. Evaluating the foundation models, LaBraM and CBraMod exhibit the strongest performance, except at 2.5\% data, where SleepFM leads (64.1\%). LaBraM and CBraMod perform similarly across most configurations, with CBraMod demonstrating a slight advantage in high-data regimes (84.39\% vs. 83.33\% at 100\% data). Compared to the supervised baseline, these architectures offer only a marginal initial improvement, such as SleepFM yielding a +0.75\% gain at 2.5\% data, before falling below the purely supervised performance in all subsequent configurations. Most importantly, our domain-specific SSL pre-training outperforms all foundation models across every data regime. SimCLR exceeds SleepFM by +6\% at 2.5\% labeled data and surpasses LaBraM by +5.39\% at the 20\% mark, while Barlow Twins outperforms CBraMod by +2.01\% using all available training data. These findings further confirm the superior effectiveness of our domain-specific SSL pre-training pipeline compared to relying on general-purpose representations.

\subsubsection{Scenario 3: SSL and Supervised Cross-Validation Within BOAS}

Finally, Table \ref{tab:res-only-std3} presents the results for scenario 3, where reported percentages of unlabeled and labeled data refer to the BOAS recordings used for SSL and supervised training, respectively. To ensure a fair comparison, equivalent proportions of HOGAR data were used for SSL pre-training. BENDR and MAEEG were excluded from this scenario due to their substantial computational demands during pre-training, although similar insights would be expected for these methods. As previously observed, the most significant improvements corresponds to low-data scenarios, where SimCLR reaches a +8.9\% and +9.98\% accuracy improvement by pre-training on the BOAS and HOGAR datasets, respectively, using only 7.5\% of labeled BOAS data. However, as the labeled data increases, the pool of available recording for SSL diminishes, reducing pre-training effectiveness. This leads to convergence toward the supervised baseline in high-data regimes and even penalizes some techniques (e.g., BYOL). Overall, pre-training on HOGAR yields performance that is comparable to or even better than using BOAS, highlighting the robustness and generalizability of representations learned from different populations and recording conditions.

\begin{table}[!t]
\renewcommand{\arraystretch}{1.3}
\setlength\tabcolsep{1.75pt}
\centering
\caption{Accuracy and standard deviation results for evaluation scenario 3: SSL and Supervised Cross-Validation Within BOAS.}
\vspace{-0.25cm}
\scriptsize
\begin{tabular}{@{}lccccc@{}}
\toprule
\multirow{2}{*}{} & \multicolumn{5}{c}{Percentage of unlabeled-labeled data} \\ \cmidrule(l){2-6} 
 & 92.5 - 7.5 & 80 - 20 & 70 - 30 & 60 - 40 & 50 - 50 \\ \midrule
\textit{Supervised} & 69.46$\pm$4.04 & 82.44$\pm$1.29 & 83.70$\pm$0.95 & 85.25$\pm$0.36 & 85.70$\pm$0.56 \\ \midrule
BOAS $\rightarrow$ BOAS &  &  &  &  &  \\ \midrule
\textit{SimCLR} & 78.36$\pm$5.63 & 83.65$\pm$0.36 & 84.82$\pm$1.43 & 84.89$\pm$0.69 & 85.34$\pm$1.22 \\
\textit{BYOL} & 67.74$\pm$7.34 & 75.55$\pm$10.93 & 82.91$\pm$0.53 & 84.00$\pm$1.15 & 84.56$\pm$1.19 \\
\textit{SimSiam} & 73.01$\pm$7.56 & 82.09$\pm$1.08 & 84.76$\pm$0.58 & 85.23$\pm$0.13 & 85.61$\pm$0.32 \\
\textit{Barlow Twins} & 77.72$\pm$5.24 & \textbf{84.57$\pm$1.00} & \textbf{85.11$\pm$0.68} & 85.10$\pm$0.12 & 85.81$\pm$0.21 \\
\textit{ContraWR} & 76.35$\pm$4.67 & 83.24$\pm$1.10 & 84.30$\pm$1.18 & 85.16$\pm$0.17 & 85.96$\pm$0.01 \\ \midrule
HOGAR $\rightarrow$ BOAS &  &  &  &  &  \\ \midrule
\textit{SimCLR} & \textbf{79.44$\pm$4.69} & 83.76$\pm$1.05 & 84.76$\pm$0.97 & 84.92$\pm$0.13 & 85.60$\pm$0.62 \\
\textit{BYOL} & 67.23$\pm$6.18 & 81.82$\pm$0.99 & 83.18$\pm$1.20 & 84.00$\pm$0.80 & 84.11$\pm$1.18 \\
\textit{SimSiam} & 76.68$\pm$5.74 & 83.27$\pm$0.62 & 84.14$\pm$1.26 & 85.14$\pm$0.03 & 85.70$\pm$1.15 \\
\textit{Barlow Twins} & 77.08$\pm$7.79 & 84.21$\pm$1.31 & 84.65$\pm$0.82 & \textbf{85.33$\pm$0.55} & \textbf{86.22$\pm$0.11} \\
\textit{ContraWR} & 77.34$\pm$4.83 & 82.84$\pm$2.24 & 84.65$\pm$0.62 & 85.17$\pm$0.53 & 85.38$\pm$1.37 \\ \bottomrule
\end{tabular}

\label{tab:res-only-std3}
\vspace{-0.2cm}
\end{table}

\subsection{Visualization of Learned Feature Representations}

\begin{figure*}[!t]
\centering
\includegraphics[width=0.95\linewidth]{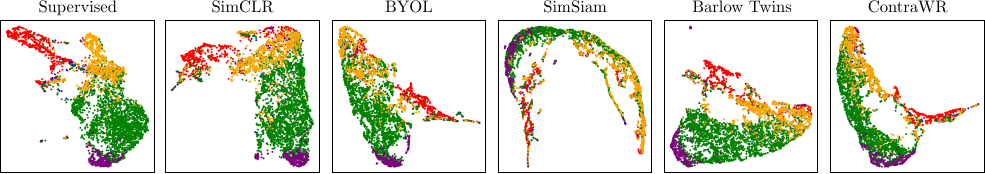}
\vspace{-0.15cm}
\caption{UMAP visualization of feature representations from the first five BOAS recordings (\protect\tikz\protect\draw[mplred,fill=mplred] (0,0) circle (.5ex); Wake, \protect\tikz\protect\draw[mplblue,fill=mplblue] (0,0) circle (.5ex); N1, \protect\tikz\protect\draw[mplgreen,fill=mplgreen] (0,0) circle (.5ex); N2, \protect\tikz\protect\draw[mplpurple,fill=mplpurple] (0,0) circle (.5ex); N3, \protect\tikz\protect\draw[mplorange,fill=mplorange] (0,0) circle (.5ex); REM). The supervised baseline was trained using BOAS labels, while SSL methods were pre-trained only on unlabeled HOGAR data without supervised fine-tuning. BENDR and MAEEG were excluded, as they produce 2D convolutional outputs rather than 1D feature representation vectors.}
\label{fig:umap}
\vspace{-0.25cm}
\end{figure*}

Fig. \ref{fig:umap} illustrates the UMAP projections of the output embeddings from the epoch encoder $f_{\theta}$. The distribution of the data points reveals overlap between the sleep stage classes, consistent with previous findings \cite{jiang2021self, mai2021bootstrapnet, yang2021self, ye2021cosleep, kumar2022muleeg}. The fully supervised baseline yields the clearest class separability, reinforcing the differences between the EEG hallmarks of each sleep phase. Among the SSL approaches, SimCLR demonstrates comparable clustering performance to that of the supervised approach. Interestingly, BYOL also shows a well-structured feature space, despite achieving lower classification accuracy. Barlow Twins and ContraWR exhibit similar visual separability, while SimSiam displays the least structured feature distribution. These results reinforce that SSL models are capable of learning structured and discriminative feature spaces without access to sleep labels.

\section{Discussion}\label{sec:discussion}

\subsection{Feature Representation and Transferability}

Our experiments demonstrate the superiority of the proposed SSL framework over the traditional supervised learning approach. Scenarios 1 and 2 show consistent improvements across different levels of labeled data, confirming that SSL pre-training is highly effective for downstream sleep scoring. Scenario 3 further validates that feature representations learned from the HOGAR dataset (inter-dataset) are as effective as those learned directly from the BOAS dataset (intra-dataset). This outcome aligns with our initial expectations, given that both databases were collected using the same device. However, the datasets differ in some aspects, such as population characteristics (elderly vs. healthy), recording conditions (self-recorded at home vs. supervised in a sleep laboratory), and signal quality (HOGAR contains more noise and artifacts). Despite these differences, our results suggest that SSL representations are transferable across domains and robust to recording variability.

When comparing the overall performance of the SSL methods, those based on contrastive frameworks that leverage negative samples (SimCLR and Barlow Twins) consistently achieve the best results. ContraWR closely follows, using a different approach by approximating a global representation as contrastive information. BENDR, which employs contrastive learning for signal reconstruction in the feature space, also shows competitive performance. BYOL and SimSiam, which do not use negative samples but instead rely on self-distillation, tend to perform worse across the evaluated scenarios, suggesting that the absence of explicit negative samples may limit their effectiveness in this context. Finally, MAEEG ranks lowest, likely due to its straightforward reconstruction in the signal space using only two additional layers. Our goal is not to assert superiority among methods, but rather to empirically benchmark a diverse range of SSL strategies. We believe that further tuning of SSL methods within our implementation could help close the current performance gap, as previous research has reported more comparable performance across approaches (e.g., SimCLR and BYOL) \cite{yang2021self, lee2024neuronet}. Regarding architectural consistency, BENDR and MAEEG pre-train on the 2D outputs of the last convolutional layer rather than on 1D pooled feature vectors (see Fig. \ref{fig:ssl_methods}). Nevertheless, the absence of the final pooling operation during pre-training does not compromise the learned representations or bias the comparative results, since the exact same convolutional layers are pre-trained in all methods, and the identical downstream architecture is subsequently initialized with these learned weights.

Our specialized SSL pre-training consistently outperforms all evaluated foundation models across every data regime. Notably, even the purely supervised baseline frequently surpasses these massive architectures, demonstrating that our pipeline extracts superior, task-relevant representations compared to general-purpose models. Given that the emergence of public foundation models challenges the necessity of specialized pre-training, these findings critically validate the relevance of domain-focused SSL for leveraging the massive unlabeled EEG datasets generated by the widespread adoption of wearable devices. Several factors can explain this performance gap. First, our curated sleep network provides inherently strong standalone performance, and the HOGAR pre-training distribution closely matches the BOAS downstream target. Second, while LaBraM and CBraMod lack sleep-specific pre-training, SleepFM's multimodal approach may translate suboptimally to our single-modality task. Furthermore, these models could struggle to adapt to the low-channel density constraints of wearable EEG. These findings align with emerging literature indicating that foundation models do not yet guarantee universal superiority, often showing only marginal improvements, with classical deep learning architectures offering competitive performance \cite{liu2026eeg, lee2025large, lee2025assessing, yang2026_steegformer}. Specifically, it has been reported that well-established networks, such as XSleepNet \cite{phan2021xsleepnet} and SleepTransformer \cite{phan2022sleeptransformer}, can match or outperform current foundation models in downstream sleep staging \cite{liu2026eeg, thapa2026multimodal, fox2025foundational, van2025deep}. Importantly, we do not claim that specialized SSL inherently supplants the broader foundation model paradigm. Rather, our results highlight its current practical advantage for our specific, low-density downstream task. Further experiments are required to fully explore this dynamic, though such investigations remain beyond the scope of this study.

UMAP projections (Fig. \ref{fig:umap}) reveal that SSL-trained models produce feature clusters consistent with the physiological transitions between sleep stages \cite{Iber2007}. The observed class overlap reflects expected biological variability and reinforces the interpretability of the learned representations. However, this inherent overlap, especially among Wake, N1, and REM stages, introduces a risk of false positives in clinical applications, as models may misclassify these physiologically similar states. Consequently, such biological ambiguity represents a fundamental source of aleatoric uncertainty. Based on the characteristics of each sleep phase, it is expected that N1 samples will cluster near Wake embeddings since N1 is considered a transition period between wakefulness and sleep. Similarly, REM samples are anticipated to appear in close proximity to Wake and N1 clusters since their EEG features resemble those seen in the awake state, though typically slower and higher in amplitude. N2 embeddings are likely to diverge from these stages, reflecting a descent in brain activity. Finally, N3 cluster, which corresponds to the deepest sleep stage, is expected to be the most distant group in the plot, revealing its characteristic low-frequency delta waves.

\subsection{Impact of Label Availability on Model Performance}

The benefits of SSL are most pronounced in low-label regimes, where supervised learning struggles due to limited training data. By learning generalizable representations from large volumes of unlabeled EEG data, SSL helps bridge this gap and mitigates the data-hungry nature of deep learning. SSL pre-trained models demonstrate the ability to reach the inter-scorer agreement range of 80--85\% \cite{danker2009interrater, rosenberg2013american, lee2022interrater}, considered the benchmark for medical-grade accuracy, using only a small fraction of labeled data. In scenario 1, Barlow Twins reaches this threshold with just 7.5\% of labeled data, while the purely supervised model remains far below it. Similarly, in scenario 2, SimCLR achieves accuracy within this range using only 10\% and 20\% of labeled data. Once performance enters the 80--85\% range, SSL improvements become less pronounced, although SSL continues to offer consistent gains over the supervised approach. Conversely, these label-efficient capabilities are not observed in the evaluated foundation models, which fail to meaningfully surpass the supervised baseline, yielding only marginal gains in the most restricted settings. While recent literature suggests that such massive architectures excel in data-rich, population-level contexts but struggle in data-scarce and subject-specific environments \cite{yang2026_steegformer}, our evaluation did not reflect this trend. Instead, the foundation models failed to demonstrate a systematic performance advantage over the supervised network across varying label availability configurations in our downstream domain.

Classification accuracies slightly above 85\% currently represent the peak performance for state-of-the-art deep learning approaches \cite{zhang2024review, phan2022automatic}. Some authors argue that neither human scorers nor machine learning models can reach 100\% accuracy due to aleatoric uncertainty \cite{van2022certainty}. Evaluating EEG pathology classification, Kiessner et al. \cite{kiessner2024reaching} found that the testing error follows a saturating power-law with both model and dataset size, and empirically observed accuracies saturate at 85\%--87\%, which corresponds to the inter-rater agreement on the clinical labels. Both SSL pre-trained and supervised models are capable of surpassing this threshold when sufficient labeled data is available. This helps explain why the gains from SSL become more limited in high-label scenarios, as performance approaches the current practical ceiling for the sleep scoring task. These observations suggest a practical threshold: SSL is most beneficial when supervised performance remains below the clinical-grade benchmark. Still, given that implementing SSL pre-training over traditional sleep staging pipelines adds relatively little overhead, its inclusion likely remains worthwhile even in high-label scenarios to surpass purely supervised performance.

\subsection{Applications and Future Directions}

In this paper, we address the challenge posed by the large volumes of unlabeled EEG data generated through the widespread adoption of wearable devices in the context of automatic sleep staging, introducing a structured benchmarking framework and a specialized SSL pipeline tailored to this domain. To leverage this unlabeled data, we evaluate SSL techniques as a pre-training step within conventional supervised learning pipelines. From a practical standpoint, SSL can be incorporated into existing sequence-to-sequence sleep staging frameworks with minimal overhead, requiring no changes to the model architecture or downstream task. As a result, SSL consistently improves classification performance over purely supervised baselines, and enhances generalization and cross-dataset transferability. In scenario 1, SSL pre-training with only 7.5\% of labeled data achieves comparable accuracy to supervised learning with 15\% of labels, representing twice the labeling effort, while using 20\% of labels with SSL results in only a \mbox{-2\%} difference compared to fully supervised training with 100\% of the dataset. This makes the development of sleep scoring systems more label-efficient and cost-effective, reducing dependence on manual annotations, which are expensive to obtain and subject to inter-scorer variability. Our findings using wearable EEG data align with prior evaluations conducted on PSG data \cite{eldele2023self}, particularly in terms of performance under varying data availability and domain generalization, further reinforcing the viability of SSL for wearable sleep staging. Moreover, our work serves as a comprehensive benchmark for wearable EEG sleep scoring and extends beyond prior studies \cite{khalique2025label} by providing the first systematic evaluation of SSL in this domain. We evaluate multiple SSL techniques across diverse training scenarios to rigorously assess label efficiency, representation quality, and cross-dataset generalization, establishing a foundational reference for future research on SSL with wearable EEG.

Based on the evidence provided, the application of SSL in sleep staging pipelines proves largely valuable when labeled data is scarce, a substantial volume of unlabeled EEG sleep data is available, and baseline performance falls below the lower bound of the inter-scorer agreement \mbox{(\textless\! 80\%)}, which we identify as the medical-grade benchmark. Translating these findings to clinical practice, domain-specific SSL pre-training may provide a more reliable and effective approach to directly applying current massive, pre-trained EEG foundation models in such label-constrained scenarios. Beyond this, SSL offers valuable opportunities in commercial consumer sleep technologies, enabling regular model updates without requiring manual scoring. Its strong generalization also makes it well-suited for distributed and federated learning setups. Additionally, SSL can support continual learning and personalized model adaptation using limited user-specific data. Importantly, the SSL techniques employed are agnostic to the domain, making them applicable to other clinical areas where labeled data is even harder to obtain. 

Current trends suggest that the near future of automatic sleep staging will consolidate around the use of affordable, wearable EEG devices equipped with a reduced number of sensors, operating in home environments with minimal supervision, typically self-administered and integrated into digital therapeutic platforms. On the modeling side, SSL-based systems are rapidly advancing toward the deployment of EEG foundation models \cite{lai2025simple, liu2026eeg, lee2025comprehensive}. Although current literature \cite{liu2026eeg, lee2025large, lee2025assessing, yang2026_steegformer} and our findings indicate that these massive architectures do not yet universally surpass domain-specific pipelines, we anticipate they will eventually fulfill their core premise: delivering state-of-the-art, general-purpose representations that transfer to downstream tasks with minimal adaptation. To achieve this, future research must address present limitations and unlock capabilities that currently remain unrealized in the EEG domain. These include demonstrating strong out-of-distribution generalization, particularly across diverse sleep disorders, varying recording hardware, and multi-center cohorts, as well as enabling optimal linear probing and robust few-shot learning. Furthermore, these architectures must establish clear scaling laws that balance parameter efficiency with performance, ensure interpretability, and seamlessly adapt to the low-channel-density constraints of real-world wearable devices. Despite these developmental hurdles, foundation models retain the potential to ultimately unlock powerful solutions for EEG analysis.

\subsection{Limitations}

This work focused on the evaluation of a single model architecture for sleep staging. Prior studies have shown that performance can vary depending on the chosen backbone network \cite{eldele2023self, lee2024neuronet}. Additionally, the relatively small size of our model may constrain its representation capacity, particularly in the SSL pre-training stage. The choice of data augmentations also influences SSL performance \cite{chen2020simple, grill2020bootstrap, jiang2021self, yang2021self}, and further refinement to our datasets could yield additional gains, although this requires manual engineering, which is a common limitation of many SSL approaches. Designing end-to-end SSL strategies that also pre-train the sequence encoder could further improve performance and represents a promising direction for future research \cite{zhang2022expert}. Moreover, although we incorporated three foundation models as baselines to contextualize and validate our specialized SSL approach, the reported results should not be interpreted as a definitive claim that domain-specific SSL is universally superior to the foundation model paradigm in sleep scoring or general EEG analysis. Further experimentation is required, and a comprehensive benchmarking of foundation models for wearable sleep staging falls outside the scope of this study.

Regarding the datasets, we achieved performance comparable to state-of-the-art automatic sleep scoring \cite{zhang2024review, phan2022automatic}, which allowed us to evaluate the system under different data availability conditions. However, we did not analyze the impact of SSL pre-training dataset size due to its limited scale. We believe that at least an order of magnitude more would be needed to draw reliable conclusions, as the effect of unlabeled data availability is not comparable to that of labeled data (see the smaller impact in Table \ref{tab:res-only-std3}). In addition, the evaluation of sleep staging was limited to the BOAS dataset, which includes labeled recordings collected in parallel with PSG in a controlled sleep laboratory to enable direct comparison between wearable EEG data and the traditional gold standard. Exploring a fully home-based scenario where experts directly annotate wearable EEG as ground truth, without relying on PSG or laboratory settings, stands as a pending aspect. 

Concerning dataset characteristics, HOGAR includes elderly adults, some with cognitive decline, while BOAS consists of a more diverse healthy population. Clinically, cognitive impairment is associated with distinct physiological sleep alterations. In terms of sleep architecture, these individuals exhibit increased sleep fragmentation, frequent awakenings, and reduced slow-wave and REM sleep \cite{reynolds1985eeg, petit2004sleep, zhang2022sleep}. At the microstructural level, EEG presents generalized slowing \cite{brayet2016quantitative, d2021eeg}, decreased coherence \cite{dunkin1994reduced}, increased theta and delta activity during wakefulness \cite{prichep1994quantitative}, reduced theta and delta in slow-wave sleep \cite{westerberg2012concurrent}, and a deterioration in non-REM sleep spindles \cite{petit2004sleep}. Although SSL demonstrated robustness in this context, these differences could still affect performance, and greater population diversity would help better capture data distribution. Extrapolating these findings to other clinical contexts requires further validation on populations with a wider age range and more varied sleep pathologies, such as insomnia or sleep apnea. Furthermore, our findings were obtained using a wearable EEG headband with which we achieved strong classification performance, supporting its validity as a suitable device for sleep staging. This allowed us to frame our discussion around achieving medical-grade performance as the objective for such systems. In this sense, the insights of this work may not generalize to other EEG hardware setups, particularly those with lower signal quality and reduced classification performance. Additionally, both datasets were recorded using the same EEG configuration (AF7 and AF8) under sleep conditions. Evaluating SSL capabilities on multimodal sources and different hardware setups remains an open aspect of this work.

\section{Conclusions}\label{sec:conclusions}

This study presents the first systematic evaluation of self-supervised learning (SSL) for sleep staging using wearable EEG, introducing a structured benchmarking framework and a specialized SSL pipeline tailored to this domain. By incorporating SSL pre-training into traditional sequence-to-sequence sleep frameworks, we address the challenge posed by the substantial volumes of unlabeled EEG data generated through the growing adoption of wearable systems. Our experiments demonstrate that SSL effectively learns generalizable feature representations from unlabeled signals and consistently improves classification performance over purely supervised baselines, particularly in low-label regimes. Moreover, SSL reaches medical-grade accuracy using under 10\% of labeled data, whereas the supervised approach requires nearly twice the labels. Remarkably, models pre-trained on unlabeled, home-recorded data generalize well to clinical datasets, highlighting the ability of SSL to produce robust and transferable representations. In addition, within the context of the evaluated datasets, the proposed specialized SSL pipeline outperforms the selected general-purpose foundation models across all data configurations. From a practical clinical perspective, our results suggest that domain-specific SSL pre-training may currently be a more appropriate strategy for data-limited, context-specific wearable sleep staging scenarios. These findings position SSL as a powerful tool for label-efficient training in wearable EEG applications, reducing reliance on costly manual annotations while enabling the democratization of sleep evaluation and potentially extending to other neurophysiological monitoring tasks. Overall, SSL bridges the gap between clinical and wearable sleep monitoring by unlocking the value of unlabeled EEG data, paving the way for accessible, accurate, and scalable sleep diagnostics.


\appendices
\section{Data Augmentation Techniques}
\label{app:data-aug}

Two sets of data augmentations are employed in this work. The first set $T_1$, inspired by \cite{yang2021self}, contains: 

\begin{itemize}
    \item \textbf{Bandpass filtering}: First-order Butterworth filter is applied using frequency intervals of (1,5) and (30,50).

    \item \textbf{Noising}: Adds high- and/or low-frequency noise. High-frequency noise is sampled from a uniform distribution, and scaled by a noise degree and the amplitude range of the original signal. Low-frequency noise is generated similarly but downsampled to 1\% of the signal length and interpolated back to match the original length.

    \item \textbf{Channel flipping}: Flips the EEG input channels.

    \item \textbf{Time shifting}: Horizontal rotation of the signal, which is split into two pieces and then resembled.
    
\end{itemize}

The second set of transformations $T_2$, adapted from \cite{jiang2021self, mai2021bootstrapnet}, includes:

\begin{itemize}
    \item \textbf{Permutation}: The signal is randomly divided into $n \in [5, 20]$  segments of unequal length, shuffled and concatenated.

    \item \textbf{Crop and resize}: A random segment of length $m \in [0.25, 0.75]$ of the original signal is cropped. Then, a linear interpolation is performed to restore the signal to its original length. 

    \item \textbf{Cutout and resize}: The signal is randomly divided into $n \in [5, 20]$ segments of unequal length, and one of them is discarded. The remaining segments are concatenated and linearly interpolated to match the original length. 

    \item \textbf{Random masking}: The signal is randomly divided into $n \in [5, 20]$ segments of unequal length. Then, a proportion of $m \in [0.25, 0.75]$ of the total segments are selected and masked with zeros. 
\end{itemize}

\section{Experimental Settings and SSL Hyperparameters}
\label{app:ssl-config}

The learning framework was executed on a Linux Ubuntu 22.04 LTS system with an Intel i9-10900K CPU at 3.70 GHz, 64 GB RAM, and an NVIDIA 3080 GPU. The algorithms were developed in Python 3.9 using PyTorch version 2.3. Table \ref{tab:ssl-config} presents the hyperparameters for each SSL method, where $\lambda_{\text{opt}}$ denotes the weight decay for the optimizer, $\lambda_{\text{loss}}$ is a weighting parameter in the Barlow Twins loss function, $\tau_{\text{loss}}$ is the temperature parameter in the loss, $\tau_{\text{ema}}$ is the target decay rate for the exponential moving average, and $\delta$ and $\sigma$ are ContraWR specific hyperparameters. These values were determined through an ablation study, employing the Adam optimizer $(lr = 0.0001,\ \beta_1 = 0.9, \beta_2 = 0.999, \epsilon = 1^{-8})$. For BENDR and MAEEG, we adopted values proposed in the respective original works, with a few exceptions: Transformer encoder depth was reduced to 4 layers, batch size was set to 64, Adam optimizer used $(lr = 0.0001,\ \lambda_{\text{opt}} = 1^{-4}, \beta_1 = 0.9,\ \beta_2 = 0.999,\ \epsilon = 1^{-8})$, and the number of epochs was set to 200.

\begin{table}[t]

\renewcommand{\thetable}{\Alph{section}.1}

\renewcommand{\arraystretch}{1.3}
\setlength\tabcolsep{3.25pt}
\scriptsize
\centering

\caption{Hyperparameter values for SSL pre-training techniques.}
\label{tab:ssl-config}

\vspace{-0.25cm}

\begin{tabular}{@{}lcccccccc@{}}
\toprule
SSL method & Epochs & Batch size & $\lambda_{\text{opt}}$ & $\lambda_{\text{loss}}$ & $\tau_{\text{loss}}$ & $\tau_{\text{ema}}$ & $\delta$ & $\sigma$ \\ \midrule
\textit{SimCLR} & 350 & 512 & $1^{-4}$ & - & 0.1 & - & - & - \\
\textit{BYOL} & 200 & 512 & $1^{-4}$ & - & - & 0.9 & - & - \\
\textit{SimSiam} & 100 & 512 & $1^{-6}$ & - & - & - & - & - \\
\textit{Barlow Twins} & 350 & 512 & $1^{-4}$ & 0.005 & - & - & - & - \\
\textit{ContraWR} & 100 & 512 & $1^{-7}$ & - & 1.0 & 0.999 & 0.1 & 2.0 \\ \bottomrule
\end{tabular}

\vspace{-0.2cm}
\end{table}

\section{Foundation Models Configuration and Adaptation}
\label{app:fm-config}

\subsection{Data Preprocessing and Tokenization}

\noindent\textit{LaBraM and CBraMod.}
Data preprocessing remains consistent with our methodology, with the exception that EEG signals are downsampled from 256 Hz to 200 Hz to align with the original implementations. Each 30-second EEG epoch, denoted as $x_i \in \mathbb{R}^{C \times T}$, is segmented into 1-second temporal patches. Given an input sequence of $L$ consecutive epochs $(x_1, \dots, x_L)$, each epoch $x_i$ is reshaped into a tensor $x'_i \in \mathbb{R}^{C \times n \times t}$, where $t = 200$ time steps per patch and the number of patches $n = \lfloor T / t \rfloor$. Since the total number of time steps per epoch is $T = 6000$ (30 s $\times$ 200 Hz), this yields $n = 30$ patches per epoch.

\vspace{4mm}

\noindent\textit{SleepFM.}
SleepFM natively accepts inputs at 128 Hz, matching our original sampling rate, thus requiring no resampling. Internally, the model segments the temporal dimension into 5-second tokens. Consequently, each 30-second epoch $x_i$ is reshaped into a tensor $x'_i \in \mathbb{R}^{C \times n \times t}$, where $t = 640$ time steps (5 s $\times$ 128 Hz) and $n = 3840 / 640 = 6$ patches per epoch.

\subsection{Model-Specific Details}

\noindent\textit{LaBraM (5.8M parameters).}
We map the input channels to the AF7 and AF8 spatial embeddings. We utilize the publicly available \texttt{LaBraM-Base} checkpoint. Because the model architecture imposes a maximum input length of 15 seconds, each 30-second epoch is split into two equal halves. The outputs of both halves are subsequently summed to produce a final 1D feature vector $h_i \in \mathbb{R}^{200}$ for the complete epoch.

\vspace{4mm}

\noindent\textit{CBraMod (4.0M parameters).}
The architecture natively processes the complete 30-second input without length constraints, yielding a 1D feature vector $h_i \in \mathbb{R}^{200}$ per epoch.

\vspace{4mm}

\noindent\textit{SleepFM (4.4M parameters).}
The network outputs a 1D feature vector $h_i \in \mathbb{R}^{128}$ representing each 30-second epoch.

\subsection{Architectural Integration}

We integrate these foundation models as backbone encoders within our sequence-to-sequence pipeline ($f_{\theta}$ in Fig. \ref{fig:pipeline}), initializing them with the pre-trained weights provided by their respective authors without conducting further pre-training. Compared to our proposed CNN epoch encoder, these models differ primarily in their architecture (utilizing large-scale Transformer backbones) and their pre-training data (leveraging much larger, heterogeneous corpora).

The 1D output feature vectors representing each epoch, denoted as $(h_1, \dots, h_L)$, are passed as input to the identical CNN sequence encoder utilized in the second stage of our proposed architecture (see Section \ref{sec:model_arq}). The downstream fine-tuning configuration for these models mirrors the setup described in Section \ref{sec:evaluation}, maintaining identical data partitions, normalization procedures, window sizes, training epochs, optimizers, and loss functions for full model fine-tuning in a semi-supervised setting.


\section*{Acknowledgements}

This work was supported by the Ministerio de Ciencia, Innovación y Universidades and the Agencia Estatal de Investigación (DIN2024-013427, MICIU/AEI/10.13039/501100011033); the Horizon-RIA-2023 call (MANOLO: 101135782); the Proyecto Estratégico para la Recuperación y Transformación Económica (PERTE) para la Salud de Vanguardia (Multi-País: EXP - 00170833 / PAIS-20241086); and the Diputación General de Aragón - Subvenciones I+D Movilidad sostenible y sector farmacéutico (ORDEN EPE/676/2023, expediente IDMF/2023/0007).


\section*{Data Availability}

The BOAS dataset is publicly available at OpenNeuro (\url{https://openneuro.org/datasets/ds005555}). The HOGAR dataset is part of an ongoing project at the date of submission and subject to ethical and regulatory constraints due to the inclusion of cognitively impaired patients. Data supporting the findings of this study can be available from the authors upon reasonable request.


\ifCLASSOPTIONcaptionsoff
  \newpage
\fi

\bibliographystyle{IEEEtran}
\bibliography{references}

\end{document}